\documentclass[12pt]{article}
\pdfoutput=1

\usepackage[utf8]{inputenc}
\usepackage[margin=1in]{geometry}
\usepackage[tbtags]{amsmath} 				
\usepackage{amssymb, mathrsfs}			
\usepackage{physics}
\usepackage{graphicx}			
\usepackage{tensor}				
\usepackage[makeroom]{cancel}	
\usepackage{bm}								
\usepackage{bbm}
\usepackage[
      colorlinks=true,
      linkcolor=blue,
      urlcolor=blue,
      filecolor=black,
      citecolor=red,
      pdfstartview=FitV,
      pdftitle={},
      pdfauthor={Matteo Vicino, Michael Gutperle},
      bookmarksopen=true
      ]{hyperref}

\usepackage{sectsty}
\sectionfont{\large}

\renewcommand{\thefootnote}{\fnsymbol{footnote}}
\renewcommand{\thanks}[1]{\footnote{#1}}
\newcommand{\starttext}{
\setcounter{footnote}{0}
\renewcommand{\thefootnote}{\arabic{footnote}}}
\renewcommand\({\begin{equation}}		
\renewcommand\){\end{equation}}
\renewcommand{\epsilon}{\varepsilon}	

\numberwithin{equation}{section} 		

\numberwithin{equation}{section}

\long\def\symbolfootnote[#1]#2{\begingroup%
\def\thefootnote{\fnsymbol{footnote}}\footnote[#1]{#2}\endgroup}

\begin{document}
\setlength{\baselineskip}{16pt}

\starttext
\setcounter{footnote}{0}

\begin{flushright}
\today
\end{flushright}

\bigskip

\begin{center}

{\Large \bf  Holographic Surface  Defects in $D=5$, $N=4$ Gauged Supergravity}

\vskip 0.4in

{\large  Michael Gutperle and Matteo Vicino}

\vskip 0.2in

{ \sl Mani L. Bhaumik Institute for Theoretical Physics} \\
{\sl Department of Physics and Astronomy }\\
{\sl University of California, Los Angeles, CA 90095, USA} 

\bigskip

\end{center}
 
\begin{abstract}
\setlength{\baselineskip}{16pt}

Solutions describing holographic surface defects in $D=5,N=4$ gauged supergravity theories are constructed. It is shown that a surface defect solution in pure Romans' gauged supergravity is singular. Adding a single vector multiplet allows for the construction of a non-singular solution. The on-shell action and one point functions of operators in the presence of the defect are computed using holographic renormalization.

\end{abstract}

\setcounter{equation}{0}
\setcounter{footnote}{0}

\newpage

\section{Introduction}

Holography is a powerful tool for studying quantum field theories.  Using holography, extended defect operators such as Wilson lines, surface operators, and domain walls can be studied.  In theories with holographic duals, there are two methods leading to the construction of the duals of $p$-dimensional defect operators in $d$-dimensional CFT. First, one can consider probe branes embedded in an $AdS_{p+1}$ slice of $AdS_{d+1}$ and in some cases wrapping some other manifold. When the number of probe branes is small, the backreaction can be neglected and the probe brane provides a good description of the defect in the dual gauge theory \cite{Karch:2000gx,DeWolfe:2001pq}. A defect will preserve some supersymmetry  if a $\kappa$-symmetry projector for the probe exists in the $AdS$ background \cite{Skenderis:2002vf}.

The second method involves searching for supergravity solutions that are warped products of an $AdS_{p+1}$ factor (together with other spaces such as spheres) over base spaces such as a Riemann manifold with boundary.  The solutions, which are often called Janus solutions \cite{Bak:2003jk}, are locally asymptotic to $AdS_{d+1}$ and describe a backreacted geometry dual to a defect. BPS solutions are obtained by imposing the vanishing of the fermionic supersymmetry transformations in a bosonic background. These BPS equations are generally easier to solve than the equations of motion.  Some examples of such solutions are Janus domain wall solutions \cite{DHoker:2007zhm,DHoker:2007hhe},  Wilson lines \cite{DHoker:2007mci} and surface operators  \cite{Lin:2004nb} in type IIB supergravity and Janus solutions in M-theory \cite{DHoker:2008lup,DHoker:2009lky}\footnote{see \cite{Yamaguchi:2006te,Gomis:2006cu,Lunin:2006xr,Lunin:2007ab,Clark:2005te} for earlier work on holographic defect solutions.}

The solutions listed above preserve sixteen of the original thirty-two supersymmetries and the large amount of supersymmetries allows for the construction of large families of exact solutions. The possibility of finding holographic duals of defect operators in supergravity backgrounds which are dual to less supersymmetric theories is an interesting question. There are large classes of $d=4, N=2$ SCFTs and several constructions of holographic duals (see e.g. \cite{Lin:2004nb,Gaiotto:2009gz,Aharony:2012tz}). These supergravity solutions are considerably more complicated than the $AdS_5\times S^5$ dual of $N=4$ SYM. Consequently, the construction of holographic duals for defects in $N=2$ SCFTs in type II or M-theory is challenging.   Instead of considering the full ten or eleven dimensional theory, it is simpler to consider a lower dimensional gauged supergravity and construct defect solutions there. In special cases, lower dimensional supergravities are consistent truncations  and solutions can be uplifted to the full ten or eleven dimensional theory. Even if this is not the case, the lower dimensional theories are still useful for studying general aspects of the defect solutions and may reveal clues for how to construct defect solutions in the full theory.

$N=4$ gauged supergravities in five dimensions have sixteen supersymmetries and their $AdS_5$ vacua can be used to describe four dimensional $N=2$ SCFTs. The pure gauged supergravity was constructed in  \cite{Romans:1985ps, Awada:1985ep}, whereas the addition of matter multiplets and general gaugings were constructed in \cite{DallAgata:2001wgl,Schon:2006kz}. The $AdS_5$ vacua and moduli spaces for these theories were analyzed in \cite{Louis:2015dca}. Some recent papers studying solutions in these theories can be found in \cite{Cassani:2012wc,Bobev:2019ylk,BenettiGenolini:2017zmu,Dao:2018xya,Dao:2018gma}.

In the present paper, we study $D=5,N=4$ gauged supergravity solutions which are dual to surface defects in the $N=2$ SCFTs.  The structure of the paper is as follows:
In Section \ref{sec2}, we briefly review the pure $D=5, N=4$ gauged supergravity of Romans. We consider an ansatz for the defect solution of the form $AdS_3\times S^1$ warped over an interval.  Such an ansatz can be related to a charged black hole by double analytic continuation and it is shown that there is no global regular solution for the defect as a conical deficit or excess in either the bulk or boundary cannot be removed. In Section \ref{sec3}, we review the matter coupled theory and its gaugings, and show that completely regular solutions can be constructed for this theory. In Section \ref{sec4}, we utilize these solutions to calculate holographic observables, namely the one point functions of operators in the presence of the defect as well as the on-shell supergravity action which is related to the free energy in the presence of the defect. We discuss the results and some directions for future research in Section \ref{sec5}. In Appendix \ref{appa}, we present details of the spin connection and the form of supersymmetry transformations used in the main part of the paper. We also show that the solution in Section \ref{sec4} preserves eight of the sixteen supersymmetries.  In Appendix \ref{appb}, we present a solution corresponding to a line defect in the Euclidean $N=4$ gauged supergravity.

\section{Romans' Gauged $N=4$ Supergravity}
\label{sec2}

The field content of Romans' gauged supergravity   \cite{Romans:1985ps, Awada:1985ep} is given by the   $N=4$ gauged supergravity multiplet
\begin{equation}
\left( e_{\mu}^{\hspace{.2cm} r}, \psi_{\mu a}, a_{\mu}, A_{\mu}^{I}, B_{\mu \nu}^{\alpha}, \chi_{a}, \phi \right)
\end{equation}
which contains the graviton $e_{\mu}^{\hspace{.2cm} r}$, four gravitini $\psi_{\mu a}$, a $U(1)$ gauge field $a_{\mu}$, an $SU(2)$ Yang-Mills gauge field $A_{\mu}^{I}$, two antisymmetric tensor fields $B_{\mu \nu}^{\alpha}$, four spin 1/2 fermions $\chi_{a}$, and a single scalar $\phi$. In the above, indices $a, b = 1, 2,3,4$ are $\text{Spin}(5) \cong USp(4)$ indices, $I, J, K = 1,2,3$ are $SU(2)$ adjoint indices, and $\alpha, \beta = 4,5$ are $SO(2) \cong U(1)$ indices. All fermionic fields satisfy the symplectic Majorana condition. We review our conventions in Appendix \ref{appa}.

The bosonic Lagrangian is given by
\begin{equation}\label{lagra}
\begin{split}
e^{-1} \mathcal{L} &= -\frac{1}{4}R -\frac{1}{4} \xi^{-4} f^{\mu \nu}f_{\mu\nu} - \frac{1}{4} \xi^2 \left(F^{\mu \nu I} F_{\mu \nu}^{I} + B^{\mu \nu \alpha} B_{\mu \nu}^{\alpha} \right) + \frac{1}{2} \partial^{\mu} \phi \partial_{\mu} \phi \\
&+ \frac{1}{4}e^{-1} \epsilon^{\mu \nu \rho \sigma \tau} \left( \frac{1}{g_{1}} \epsilon_{\alpha \beta} B_{\mu \nu}^{\alpha} D_{\rho} B_{\sigma \tau}^{\beta} - F_{\mu \nu}^{I} F_{\rho \sigma}^{I} a_{\tau}\right) + V(\phi)
\end{split}
\end{equation}
where the field strengths and scalar potential take the form
\begin{equation}
\begin{split}
& f_{\mu \nu} = \partial_{\mu} a_{\nu} - \partial_{\nu} a_{\mu} \\
& F_{\mu \nu}^{I} = \partial_{\mu} A_{\nu}^{I} - \partial_{\nu} A_{\mu}^{I} + g_{2} \epsilon^{IJK} A_{\mu}^{J} A_{\nu}^{K} \\
& V = \frac{g_{2}}{8} \left(g_{2} \xi^{-2} + 2\sqrt{2} g_{1} \xi \right) \\
& \xi = \text{exp}\left( \sqrt{\frac{2}{3}} \phi \right)
\end{split}
\end{equation}
This Lagrangian (\ref{lagra}) leads to the equations of motion
	\begin{equation}\label{equam}
	\begin{split}
	&R_{\mu \nu} - 2 \partial_{\mu} \phi \partial_{\nu} \phi -\frac{4}{3} V(\phi) g_{\mu \nu} + \xi^{-4} \left( 2f_{\mu \rho} f_{\nu}^{\hspace{.2cm} \rho} - \frac{1}{3} g_{\mu \nu} f_{\rho \sigma} f^{\rho \sigma} \right) \\
	& \hspace{.75cm} + \xi^{2} \left( 2F_{\mu \rho}^{I} F_{\nu}^{I \hspace{.1cm} \rho} + 2B_{\mu \rho}^{\alpha} B_{\nu}^{\alpha \hspace{.1cm} \rho} - \frac{1}{3} g_{\mu \nu} \left( F_{\rho \sigma}^{I} F^{I \rho \sigma} + B_{\rho \sigma}^{\alpha} B^{\rho \sigma \alpha}  \right) \right)  = 0 \\
	&-\Box \phi + \frac{\partial V}{\partial \phi} + \sqrt{\frac{2}{3}} \xi^{-4} f_{\mu \nu} f^{\mu \nu} -\frac{1}{\sqrt{6}}  \left( F_{\mu \nu}^{I} F^{I \mu \nu} + B_{\mu \nu}^{\alpha} B^{\mu \nu \alpha}  \right) = 0 \\
	&D_{\nu} \left( \xi^{-4} f^{\nu \mu} \right) -\frac{1}{4} e^{-1} \epsilon^{\mu \nu \rho \sigma \tau}  \left( F_{\nu \rho}^{I} F^{I}_{ \sigma \tau} + B_{\nu \rho}^{\alpha} B^{\alpha}_{\sigma \tau}  \right) = 0 \\
	& D_{\nu} \left(\xi^{2} F^{\nu \mu I} \right) - \frac{1}{2} e^{-1} \epsilon^{\mu \nu \rho \sigma \tau} F^{I}_{\nu \rho} f_{\sigma \tau} = 0 \\
	& e^{-1} \epsilon^{\mu \nu \rho \sigma \tau} \epsilon^{\alpha \beta} \mathcal{D}_{\rho} B_{\sigma \tau} ^{\beta} - g_{1} \xi^2 B^{\alpha \mu \nu} = 0
	\end{split}
	\end{equation}
where the covariant derivative acting on a vector representation is
\begin{equation}
D_{\mu} V^{I \alpha} = \nabla_{\mu} V^{I \alpha} + g_{1} a_{\mu} \epsilon^{\alpha \beta} V^{I \beta} + g_{2} \epsilon^{IJK} A_{\mu}^{J} V^{K \alpha}
\end{equation}
The supersymmetry transformation of the fermions are
	\begin{equation}
	\begin{split}
	& \delta \psi_{\mu a} = D_{\mu}\epsilon_{a} + \gamma_{\mu} T_{ab} \epsilon^{b} - \frac{1}{6\sqrt{2}} ({\gamma_{\mu}}^{\nu \rho} - 4 \delta_{\mu}^{\nu} \gamma^{\rho} )\left(H_{\nu \rho a b} + \frac{1}{\sqrt{2}} h_{\nu \rho a b} \right) \epsilon^{b} \\
	& \delta \chi_{a} = \frac{1}{\sqrt{2}} \gamma^{\mu} \partial_{\mu} \phi \epsilon_{a} + A_{ab} \epsilon^{b} - \frac{1}{2\sqrt{6}} \gamma^{\mu \nu}(H_{\mu \nu a b} - \sqrt{2} h_{\mu \nu a b}) \epsilon^{b}
	\end{split}
	\end{equation}
where the action of the covariant derivative on a spinor is
	\begin{equation}
	 D_{\mu} \epsilon_{a} = \nabla_{\mu} \epsilon_{a} + \frac{1}{2} g_{1} a_{\mu} (\Gamma_{45})_{a}^{\hspace{.2cm}b} \epsilon_{b} + \frac{1}{2} g_{2} A_{\mu}^{I} (\Gamma_{I45})_{a}^{\hspace{.2cm} b} \epsilon_{b}
	\end{equation}
and
	\begin{equation}
	\begin{split}
	& H_{\mu \nu}^{ab} = \xi ( F_{\mu \nu}^{I} (\Gamma_{I})^{ab} + B_{\mu \nu}^{\alpha} (\Gamma_{\alpha})^{ab} ) \\
	& h_{\mu \nu}^{ab} = \xi^{-2} \Omega^{ab} f_{\mu \nu} \\
	& T^{ab} = \frac{1}{6}\left( \frac{1}{\sqrt{2}} g_{2} \xi^{-1} + \frac{1}{2} g_{1} \xi^2 \right) (\Gamma_{45})^{ab} \\
	& A^{ab} = \frac{1}{2\sqrt{3}} \left( \frac{1}{\sqrt{2}} g_{2} \xi^{-1} - g_{1} \xi^2 \right) (\Gamma_{45})^{ab}
	\end{split}
	\end{equation}
The matrices $\Gamma_{i}$ satisfy the $D=5$ Euclidean Clifford algebra
	\begin{equation}
	\left( \Gamma_{i} \right)_{a}^{\hspace{.2cm}b} \left( \Gamma_{j} \right)_{b}^{\hspace{.2cm}c} + \left( \Gamma_{j} \right)_{a}^{\hspace{.2cm}b} \left( \Gamma_{i} \right)_{b}^{\hspace{.2cm}c}= 2 \delta_{ij} \delta_{a}^{c}
	\end{equation}
and the charge conjugation matrix $\Omega^{ab} = - \Omega^{ba} $ can be used to raise or lower spinor indices
	\begin{equation}
	\epsilon^{a} = \Omega^{ab} \epsilon_{b} \hspace{1cm} \epsilon_{a} = \Omega_{ab} \epsilon^{b}
	\end{equation}
so that $\Omega_{ab} \Omega^{bc} = \delta_{a}^{c}$ for consistency. $\Gamma_{5}$ is chosen such that $\left(\Gamma_{12345} \right)_{a}^{\hspace{.2cm}b} = \delta_{a}^{b}$. As discussed in \cite{Romans:1985ps}, different choices of the parameters $g_1$ and $g_2$ correspond to different gauged supergravities.  For the choice  $g_{2} = \sqrt{2} g_{1} = 2\sqrt{2}$, the theory has an Anti-de Sitter vacuum with radius of curvature $L_{AdS}=1$ and preserves  sixteen supersymmetries. These values of the couplings are used in what follows.
The bosonic and fermionic supersymmetries combine into the superalgebra $SU(2,2|2)$ which is also the superconformal algebra of $d=4,N=2$ SCFTs. 

\subsection{Half-BPS Surface Defect  in Romans' Theory}
The superalgebra  $SU(2,2|2)$ contains a superalgebra  $SU(1,1|1) \times SU(1,1|1) \times U(1)$, which has eight odd generators and an even  $SO(2,1)\times SO(2,1)\times U(1)^3 \cong SO(2,2) \times U(1)^3$ subalgebra. Such an unbroken superalgebra corresponds to half-BPS superconformal surface operators in $N=2, d=4$ SCFTs \cite{Gaiotto:2009fs}. The even part of the subgroup can be realized holographically by the ansatz
\begin{equation}
\begin{split}
& ds^2 = f_{1}(r)^2 ds_{AdS_{3}}^2 - f_{2}(r)^2 d\theta^2 - f_{3}(r)^2 dr^2 \\
& A^{I} = \delta^{I3} A(r) d\theta
\end{split}
\end{equation}
A solution of this form can be generated by performing a double Wick rotation of the BPS black hole solution \cite{Behrndt:1998jd,Behrndt:1998ns} used in \cite{Crossley:2014oea} to calculate Super-Renyi entropies. The solution to the equations of motion is then given by
	\begin{equation}\label{1chargesol}
	\begin{split}
	& ds^{2} = r^2 H(r)^{2/3} \left(\cosh^{2}{\rho} \hspace{.1cm} dt^2 - d\rho^2 - \sinh^{2}{\rho} \hspace{.1cm} d\varphi^{2} \right) - \frac{f(r)}{H(r)^{4/3}} d\theta^2 - \frac{H(r)^{2/3}}{f(r)} dr^2 \\
		& H = 1+ \frac{q}{r^2} \hspace{1cm} f = r^2 H^2 - 1 \\
	& \xi = H^{1/3} \hspace{1cm} A^{I} = \delta^{I3} \left(\mu - \frac{q}{\sqrt{2} (r^2 + q)} \right) d\theta
	\end{split}
	\end{equation}  
This solution preserves eight of the original sixteen supersymmetries of the $AdS_5$ vacuum of Romans' theory and is a special case of the matter coupled solution that is presented in the following section. The number of supersymmetries and the verification of the equations of motion follow from the more general case considered there. 

The minimal value of the radial coordinate $r_{0}$ is determined by the largest root of $f(r)$ which previously corresponded to the outer horizon of the BPS black hole. Expanding about the origin $r_{0}$ leads to 
\begin{equation}
\begin{split}
& ds^2 \sim d\tilde{r}^2 + \left(1 - 4q \right) \tilde{r}^2 d\theta^2 \\ 
& \tilde{r} = r - r_{0} = r - \frac{1}{2} \left(1 + \sqrt{1-4q} \right)
\end{split}
\end{equation}
The boundary metric is conformal to flat space
\begin{equation}
ds^2_{\partial} = \cosh^{2}{\rho} \hspace{.1cm} dt^2 - d\rho^2 - \sinh^{2}{\rho} \hspace{.1cm} d\varphi^{2} - d\theta^2 = ds^2_{AdS_{3}} - d\theta^2
\end{equation}
which implies that there will be an angular deficit or excess in either the bulk metric or the boundary metric unless $q=0$. Regularity at the origin can be restored by coupling vector multiplets.

\section{Matter Coupled Theory}
\label{sec3}

It is possible to add matter multiplets to the pure Romans' theory.  The $N=4$ vector multiplet
\begin{equation}
\left( A_{\mu}, \lambda_{i}, \phi^{m} \right)
\end{equation} 
contains a vector field $A_{\mu}$, four fermions $\lambda_{i}$, and five scalars $\phi^{m}$. The indices $i=1,\dots,4$ and $m = 1, \dots, 5$ are $USp(4)$ and $SO(5)$ indices respectively. The matter couplings and gaugings  are completely determined  in terms of embedding tensors $\xi_{MN}$ and $f_{MNP}$ \cite{DallAgata:2001wgl,Schon:2006kz}.  The supersymmetric vacua of such theories where investigated in \cite{Louis:2015dca}.

These embedding tensors satisfy the quadratic constraints
\begin{equation}
f_{R[MN}f_{PQ]}^{\hspace{.5cm}R}=0 \hspace{1cm} \xi_{M}^{\hspace{.2cm}Q}f_{QNP} = 0
\end{equation}
and determine the gauging of the R-symmetry. It is convenient to introduce a composite index $\mathcal{M} = \{0,M\}$ such that the covariant derivative acting on a vector representation is given by
\begin{equation}
\begin{split}
 D_{\mu} V^{\mathcal{M}} &= \nabla_{\mu} V^{\mathcal{M}} + g A_{\mu}^{\mathcal{N}} X_{\mathcal{N} \mathcal{P}}^{\hspace{.5cm} \mathcal{M}} V^{\mathcal{P}} \\
X_{MN}^{\hspace{.5cm}P} &= -f_{MN}^{\hspace{.5cm}P} \hspace{.75cm} X_{0M}^{\hspace{.5cm}N} = -\xi_{M}^{\hspace{.2cm}N}
\end{split}
\end{equation}
The coupling of $n$ vector multiplets is described by a coset representative $\mathcal{V}$ of \newline $SO(5,n)/SO(5)\times SO(n)$. The coset representative $\mathcal{V}$ decomposes as
\begin{equation}
\mathcal{V} = \left( \mathcal{V}_{M}^{\hspace{.3cm}m}, \mathcal{V}_{m}^{\hspace{.3cm}a} \right)
\end{equation}
where $m=1,\dots,5$ and $a=1,\dots,n$ are $SO(5)$ and $SO(n)$ indices respectively. As an element of $SO(5,n)$, $\mathcal{V}$ must satisfy
\begin{equation}
\eta_{MN} = \mathcal{V}_{M}^{\hspace{.3cm}P} \eta_{PQ} \mathcal{V}_{N}^{\hspace{.3cm}Q} =  -\mathcal{V}_{M}^{\hspace{.3cm}m}\mathcal{V}_{N}^{\hspace{.3cm}m} + \mathcal{V}_{M}^{\hspace{.3cm}a}\mathcal{V}_{M}^{\hspace{.3cm}a}
\end{equation}
where $\eta_{MN} = \text{diag}(-1,-1,-1,-1,-1,+1,\dots,+1) $. The scalar kinetic terms are expressed in terms of the matrix
\begin{equation}
M_{MN} = \mathcal{V}_{M}^{\hspace{.3cm}m}\mathcal{V}_{N}^{\hspace{.3cm}m} + \mathcal{V}_{M}^{\hspace{.3cm}a}\mathcal{V}_{M}^{\hspace{.3cm}a}
\end{equation}
and the  bosonic Lagrangian is given by
\begin{equation}
\begin{split}
e^{-1}\mathcal{L} = & \frac{1}{2}R - \frac{1}{4} \Sigma^{2} M_{MN}\mathcal{H}_{\mu \nu}^{M} \mathcal{H}^{N \mu \nu} - \frac{1}{4}\Sigma^{-4} \mathcal{H}_{\mu \nu}^{0} \mathcal{H}^{0\mu \nu} \\
& - \frac{3}{2} \Sigma^{2} \left( \partial_{\mu} \Sigma \right)^2 + \frac{1}{16} \left( D_{\mu}M_{MN}\right) \left(D^{\mu} M^{MN} \right) -g^2 V + e^{-1}\mathcal{L}_{\text{top}}
\end{split}
\end{equation}
where $\mathcal{L}_{\text{top}}$ is a topological term. The covariant field strengths are
\begin{equation}
\begin{split}
& \mathcal{H}_{\mu\nu}^{\mathcal{M}} =\partial_{\mu}A_{\nu}^{\mathcal{M}} - \partial_{\nu} A^{\mathcal{M}}_{\mu}+ gX_{\mathcal{N} \mathcal{P}}^{\hspace{.5cm} \mathcal{M}} A_{\mu}^{\mathcal{N}} A_{\nu}^{\mathcal{P}} + g Z^{\mathcal{M} \mathcal{N} } B_{\mu \nu \mathcal{N}} \\
& Z^{MN} = \frac{1}{2}\xi^{MN}
\end{split}
\end{equation}
where $B_{\mu \nu \mathcal{M}}$ are two-form fields that are introduced in the process of gauging the theory. The scalar potential is
\begin{equation}
\begin{split}
& V = V_{1} + V_{2}  + V_{3}\\
& V_{1} = \frac{1}{4}  f_{MNP} f_{QRS} \Sigma^{-2} \left( \frac{1}{12}M^{MQ}M^{NR}M^{PS} - \frac{1}{4}M^{MQ}\eta^{NR}\eta^{PS} + \frac{1}{6}\eta^{MQ}\eta^{NR}\eta^{PS} \right) \\
& V_{2} = \frac{1}{16}\xi_{MN} \xi_{PQ} \Sigma^{4} \left( M^{MP} M^{NQ} - \eta^{MP} \eta^{NQ} \right) \\
& V_{3} = \frac{1}{6\sqrt{2}} f_{MNP}\xi_{QR} \Sigma M^{MNPQR}
\end{split}
\end{equation}
with the completely antisymmetric matrix $M_{MNPQR}$ taking the form
\begin{equation}
M_{MNPQR} = \epsilon_{mnopq}\mathcal{V}_{M}^{\hspace{.3cm}m} \mathcal{V}_{N}^{\hspace{.3cm}n} \mathcal{V}_{P}^{\hspace{.3cm}o} \mathcal{V}_{Q}^{\hspace{.3cm}p} \mathcal{V}_{R}^{\hspace{.3cm}q}
\end{equation}
The $SO(5)$ index $M$ of $\mathcal{V}_{M}$ can be converted to a pair of antisymmetric $USp(4)$ indices $ij$ through the formulas
\begin{equation}
\mathcal{V}_{M}^{\hspace{.3cm}ij} = \frac{1}{2} \mathcal{V}_{M}^{\hspace{.3cm}m} \Gamma_{m}^{ij} \hspace{1cm}
\mathcal{V}_{ij}^{\hspace{.3cm}M} = \frac{1}{2} \mathcal{V}_{m}^{\hspace{.3cm}M} \Gamma_{m}^{kl}\Omega_{ki} \Omega_{lj}
\end{equation}
with a sum over $m$. The matrices
\begin{equation}
\begin{split}
&\zeta^{ij} = \sqrt{2}\Sigma^{2} \Omega_{kl} \mathcal{V}_{M}^{\hspace{.3cm}{ik}} \mathcal{V}_{N}^{\hspace{.3cm}{jl}} \xi^{MN} \\
&\zeta^{aij} = \Sigma^{2}  \mathcal{V}_{M}^{\hspace{.3cm}{a}} \mathcal{V}_{N}^{\hspace{.3cm}{ij}}\xi^{MN} \\
&\rho^{ij}=-\frac{2}{3}\Sigma^{-1} \mathcal{V}_{M}^{\hspace{.3cm}{ik}} \mathcal{V}_{N}^{\hspace{.3cm}{jl}} \mathcal{V}^{P}_{\hspace{.3cm}{kl}}f^{MN}_{\hspace{.6cm}P}\\
&\rho^{aij} = \sqrt{2} \Sigma^{-1} \Omega_{kl} \mathcal{V}_{M}^{\hspace{.3cm}{a}} \mathcal{V}_{N}^{\hspace{.3cm}{ik}}\mathcal{V}_{P}^{\hspace{.3cm}{jl}}f^{MNP}
\end{split}
\end{equation}
appear in the fermion shift matrices
\begin{equation}
\begin{split}
& A_{1}^{ij} = \frac{1}{\sqrt{6}} \left( -\zeta^{ij} + 2\rho^{ij} \right) \\
& A_{2}^{ij} = -\frac{1}{\sqrt{6}} \left( \zeta^{ij} + \rho^{ij} \right) \\
& A_{2}^{aij} = \frac{1}{2} \left( -\zeta^{aij} + \rho^{aij} \right)
\end{split}
\end{equation}
A minus sign has been inserted into $A_{2}^{ij}$ relative to \cite{Schon:2006kz} to match the BPS equations of Romans' supergravity in a mostly plus signature as in \cite{DallAgata:2001wgl}. The BPS equations are
\begin{equation}
\begin{split}
& \delta \psi_{\mu i} = D_{\mu} \epsilon_{i} - \frac{i}{6} \left( \Omega_{ij} \Sigma \mathcal{V}_{M}^{\hspace{.3cm}{ik}} \mathcal{H}_{\nu \rho}^{M} -\frac{1}{2\sqrt{2}} \delta_{i}^{k} \Sigma^{-2} \mathcal{H}_{\nu \rho}^{0} \right)\left(\gamma_{\mu}^{\hspace{.2cm}\nu \rho} - 4\delta_{\mu}^{\nu} \gamma^{\rho}\right) \epsilon_{k} \\
& \hspace{1cm} + \frac{ig}{\sqrt{6}} \Omega_{ij} A_{1}^{jk} \gamma_{\mu} \epsilon_{k} \\
& \delta \chi_{i} = -i\frac{\sqrt{3}}{2} \left( \Sigma^{-1} \partial_{\mu} \Sigma \right)\gamma^{\mu} \epsilon_{i} - \frac{1}{2\sqrt{3}} \left( \Sigma \Omega_{ij} \mathcal{V}_{M}^{\hspace{.3cm}{jk}} \mathcal{H}^{M}_{\mu \nu} + \frac{1}{\sqrt{2}} \Sigma^{-2} \delta_{i}^{k} \mathcal{H}_{\mu \nu}^{0} \right) \gamma^{\mu \nu} \epsilon_{k} \\
& \hspace{1cm} + \sqrt{2} g \Omega_{ij}A_{2}^{kj}\epsilon_{k} \\
& \delta \lambda_{i}^{a} = i \Omega^{jk} \left( \mathcal{V}_{M}^{\hspace{.3cm}{a}} D_{\mu} \mathcal{V}_{ij}^{\hspace{.2cm}{M}} \right) \gamma^{\mu} \epsilon_{k} - \frac{1}{4}\Sigma \mathcal{V}_{M}^{\hspace{.3cm}a} \mathcal{H}_{\mu \nu}^{M} \gamma^{\mu \nu} \epsilon_{i} + \sqrt{2} g \Omega_{ij} A_{2}^{akj}\epsilon_{k}
\end{split}
\end{equation}
with the action of the covariant derivative on a spinor given by
\begin{equation}
D_{\mu} \epsilon_{i} = \nabla_{\mu} \epsilon_{i} - \mathcal{V}^{M}_{\hspace{.3cm}ik} \partial_{\mu} \mathcal{V}_{M}^{\hspace{.3cm}kj} - g A_{\mu}^{0} \xi^{MN} \mathcal{V}_{Mik} \mathcal{V}_{N}^{\hspace{.3cm}kj} + g A_{\mu}^{M} f_{MNP} \mathcal{V}^{N}_{ik} \mathcal{V}^{Pkj} 
\end{equation}

\subsection{Half-BPS Surface Defect in the Matter Coupled Theory}
The gauging corresponding to Romans' supergravity with $L_{AdS}=1$ is given by
\begin{equation}
\begin{split}
&f_{MNP} = -\frac{1}{\sqrt{2}} \epsilon_{MNP} \hspace{.5cm} M,N,P \in \{1,2,3\} \\
&\xi_{MN} = -\frac{1}{2} \left( \delta_{M}^{4} \delta_{N}^{5} - \delta_{N}^{4} \delta_{M}^{5} \right)
\end{split}
\end{equation}
We will couple one vector multiplet and choose the coset element
\begin{equation}
\mathcal{V} = \text{exp}(\phi_{3} Y_{3})
\end{equation}
with the non-compact generator $\left( Y_{3} \right)_{mn} = \delta_{3m} \delta_{6n} + \delta_{3n} \delta_{6m}$. The scalar $\phi_{3}$ is a singlet under gauge transformations generated by $\sigma_{3} \in su(2) $. The theory can be truncated to $\Sigma, \phi_{3}, A^{3}_{\mu}, A^{6}_{\mu }, g_{\mu \nu}$ and the Lagrangian is
\begin{equation}
\begin{split}
e^{-1} \mathcal{L} &= \frac{1}{2} R - \frac{1}{4} \Sigma^{2} \left[\frac{1}{2} e^{2\phi_{3}} \left(F_{\mu\nu}^{3} + F_{\mu\nu}^{6} \right)^2 + \frac{1}{2} e^{-2\phi_{3}} \left(F_{\mu\nu}^{3} - F_{\mu\nu}^{6} \right)^2 \right] \\
& -\frac{3}{2}\Sigma^{-2} \left(\partial_{\mu} \Sigma\right)^{2} - \frac{1}{2} \left( \partial_{\mu} \phi_{3} \right)^{2} + 2 \left(\Sigma^{-2} + \Sigma \left(e^{\phi_{3}} + e^{-\phi_{3}} \right) \right)
\end{split}
\end{equation}
where $A^{6}_{\mu}=A_{\mu}$ is the vector from the vector multiplet. For $\phi_{3}=A_{\mu}^{6}=0$, we recover Romans' theory with the gauge field $A_{\mu}^{3}$ rescaled. The STU model \cite{Behrndt:1998jd} can be embedded into the matter coupled theory with the identifications
\begin{equation}
\begin{split}
&T = \frac{1}{\Sigma} e^{-\phi_{3}} \\
&U = \frac{1}{\Sigma} e^{\phi_{3}} \\
& F_{\mu \nu} = F^{3}_{\mu\nu} + F^{6}_{\mu\nu} \\
& G_{\mu \nu} = F^{3}_{\mu\nu} - F^{6}_{\mu\nu} 
\end{split}
\end{equation}
The equations of motion are
\begin{equation}
\begin{split}
& R_{\mu\nu} + \frac{1}{2}\Sigma^{2}\left(e^{2\phi_{3}} F_{\mu}^{\hspace{.1cm}\alpha} F_{\alpha \nu} + e^{-2\phi_{3}} G_{\mu}^{\hspace{.1cm}\alpha} G_{\alpha \nu}\right) - 3\Sigma^{-2} \partial_{\mu}\Sigma \partial_{\nu} \Sigma - \partial_{\mu} \phi_{3} \partial_{\nu} \phi_{3} \\
& \hspace{1cm} + g_{\mu\nu} \left( \frac{1}{12} \Sigma^{2} \left( e^{2\phi_{3}} F^{\alpha \beta} F_{\alpha \beta} + e^{-2\phi_{3}} G^{\alpha \beta} G_{\alpha \beta} \right) + \frac{4}{3} \left( \Sigma^{-2} + \Sigma \left(e^{\phi_{3}} + e^{-\phi_{3}} \right) \right) \right) = 0 \\
& \frac{1}{\sqrt{-g}} \partial_{\mu} \left( \sqrt{-g}\Sigma^{-2} \partial^{\mu} \Sigma \right) +\Sigma^{-3} \left( \partial_{\mu} \Sigma \right)^2 - \frac{1}{12} \Sigma \left( e^{2\phi_{3}} F^{\mu \nu} F_{\mu \nu} + e^{-2\phi_{3}} G^{\mu \nu} G_{\mu \nu} \right) \\
& \hspace{1cm} + \frac{2}{3} \left( e^{\phi_{3}} + e^{-\phi_{3}} -2\Sigma^{-3}  \right) = 0 \\
& \frac{1}{\sqrt{-g}} \partial_{\mu} \left(\sqrt{-g} \partial^{\mu} \phi_{3} \right) - \frac{1}{4} \Sigma^{2} \left( e^{2\phi_{3}} F^{\mu \nu} F_{\mu \nu} - e^{-2\phi_{3}} G^{\mu \nu} G_{\mu \nu} \right) + 2 \Sigma \left(e^{\phi_{3}} - e^{-\phi_{3}} \right) = 0 \\
& \frac{1}{\sqrt{-g}} \partial_{\mu} \left( \sqrt{-g} \Sigma^2 e^{2\phi_{3}} F^{\mu\nu} \right) = 0 \\
& \frac{1}{\sqrt{-g}} \partial_{\mu} \left( \sqrt{-g} \Sigma^2 e^{-2\phi_{3}} G^{\mu\nu} \right) = 0
\end{split}
\end{equation}
It is straightforward to verify that the equations are solved by the double Wick rotated two charge solution of \cite{Behrndt:1998jd}
\begin{equation}\label{2chargesol}
\begin{split}
& ds^{2} = r^{2} (H_{1} H_{2})^{1/3} \left(-\cosh^{2}{\rho} \hspace{.1cm} dt^2 + d\rho^2 + \sinh^{2}{\rho} \hspace{.1cm} d\varphi^{2} \right) + \frac{f}{(H_{1}H_{2})^{2/3}} d\theta^{2} + \frac{(H_{1}H_{2})^{1/3}}{f} dr^2 \\
& H_{1} = 1 + \frac{Q}{r^2} \hspace{1.5cm} H_{2} = 1 + \frac{q}{r^2} \hspace{1cm} f = r^2 H_{1} H_{2} - 1\\
&\Sigma = (H_{1} H_{2})^{1/6} \hspace{1.5cm} e^{2\phi_{3}} = \frac{H_{1}}{H_{2}} \\
& A^{3} + A^{6} =  \left(\mu_{3} + \mu_{6} - \frac{Q}{r^2+Q} \right) d\theta \hspace{1cm} A^{3} - A^{6} = \left( \mu_{3} - \mu_{6} - \frac{q}{r^2+q} \right) d\theta
\end{split}\end{equation}
where $\mu_{3}$ and $\mu_{6}$ are the chemical potentials for $A^{3}$ and $A^{6}$ respectively. For $Q=q$ and $\mu_{6} = 0$, this solution (\ref{2chargesol}) reduces to that of the previous section (\ref{1chargesol}) upon identifying $A_{\text{new}} = \sqrt{2} A_{\text{old}}$. As before, the spacetime closes at the largest root $r_{0}$ of $f(r)$ which is now given by
\begin{equation}
r_{0}^2 = \frac{1-q-Q}{2} + \frac{1}{2} \sqrt{ 1 + (Q-q)^2 - 2(Q+q) }
\end{equation}
After expanding the bulk metric about $r_{0}$, the absence of an angular deficit or excess in both the bulk metric and the boundary metric requires
\begin{equation}
(Q-q)^2 = 2 (Q + q)
\end{equation}
It is convenient to redefine the integration constants $q$ and $Q$ as
\begin{equation}
\begin{split}
& Q = q_{1} + q_{2} \\
& q = q_{1} - q_{2}
\end{split}
\end{equation}
so that regularity at the origin requires $q_{1} = q_{2}^2$ and the spacetime closes at $r_{0}^2 = 1- q_{2}^2$. The spacetime develops a singularity at $r = 0$, but this value will be excluded from the physical range of the radial coordinate for $q_{2}^2 \leq 1$. 

In the solution (\ref{2chargesol}), both scalars have a non-trivial profile. The dilaton $\Sigma$ is regular at the origin, but the additional scalar $\phi_{3}$ contains a kink
\begin{equation}
\Sigma'(r_{0}) = 0 \hspace{1cm} \phi_{3}'(r_{0}) \neq 0
\end{equation}
For generic chemical potentials, the gauge fields have a non-zero holonomy around $r=r_{0}$. We show in Appendix \ref{appb} that the bosonic background (\ref{2chargesol}) preserves eight of the sixteen supersymmertries of the gauged supergravity.

\section{Holographic Observables}
\label{sec4}
In this section, we use holographic renormalization \cite{deHaro:2000vlm,Skenderis:2002wp} to calculate some holographic observables, namely the free energy and vacuum expectation values of operators in the presence of a surface defect.
\subsection{Free Energy}
Using the equations of motion, the on-shell action takes the form
\begin{equation*}
S_{\text{bulk}} = -\int_{\mathcal{M}} d^{5} x \sqrt{-g} \left( \frac{1}{12}\Sigma^2\left( e^{2\phi_{3}} F^{\mu \nu} F_{\mu \nu} + e^{-2\phi_{3}} G^{\mu \nu} G_{\mu \nu} \right) +\frac{4}{3} \left( \Sigma^{-2} + \Sigma \left( e^{\phi_{3}} + e^{-\phi_{3}} \right) \right) \right)
\end{equation*}
The bulk action is divergent and can be renormalized by imposing a cutoff on the spacetime. In Fefferman-Graham coordinates
\begin{equation}
ds^{2} = \frac{dz^2}{z^2} + \frac{1}{z^2} g_{ij}dx^{i} dx^{j}
\end{equation}
one imposes the cutoff $z=\epsilon$ and adds boundary counterterms. Since the regularized spacetime contains a boundary, the Gibbons-Hawking term
\begin{equation}
S_{GH} = \int_{\partial \mathcal{M}} d^{4} x \sqrt{-h} K =  - \int_{\partial \mathcal{M}} d^{4}x z \partial_{z} \sqrt{-h}
\end{equation}
must be included to maintain the variational principle of the metric. In the above formula, $h_{\mu \nu}$ is the induced metric on the boundary and $K$ is the trace of the extrinsic curvature. In the notation of \cite{BenettiGenolini:2017zmu}, the bulk fields are expanded as

\begin{equation}
\begin{split}
& g_{ij} = g^{(0)}_{ij} + z^2 g^{(2)}_{ij} + z^{4} \left( g^{(4)}_{ij} + \left(\log{z}\right)^2 h^{(0)}_{ij} + \log{z} \hspace{.1cm} h^{(1)}_{ij}\right) + \dots  \\
& \Sigma = 1 + z^{2} \left(b_{1} \log{z} + b_{2} \right) + \dots \\
& \phi_{3} = z^{2} \left(c_{1} \log{z} + c_{2} \right) + \dots \\
& F =  d \left(A_{1} + A_{2} z^2 + A_{3} z^2 \log{z} + \dots \right)\\
& G = d \left( a_{1} + a_{2} z^2  + a_{3} z^2 \log{z} + \dots \right)
\end{split}
\end{equation}
and the equations of motion are solved order by order in $z$. The expansion of the Ricci tensor is
\begin{equation}
\begin{split}
& R_{zz} = -\frac{4}{z^2} - \frac{1}{2} \text{Tr}\left[g^{-1} g'' \right] + \frac{1}{2z}\text{Tr}\left[g^{-1} g' \right] + \frac{1}{4}\text{Tr}\left[g^{-1} g' g^{-1} g' \right] \\
& R_{ij} = -\frac{4}{z^2} g_{ij} - \frac{1}{2} g''_{ij} + \frac{3}{2z} g'_{ij}+ \frac{1}{2} \left(g'g^{-1}g'\right)_{ij} - \frac{1}{4}\text{Tr}\left[g^{-1}g'\right]g'_{ij} \\
& \hspace{1.5cm}+ R[g]_{ij} + \frac{1}{2z} \text{Tr}\left[g^{-1} g' \right] g_{ij}
\end{split}
\end{equation}
where $R[g]_{ij}$ is the boundary Ricci tensor and primes denote derivatives with respect to $z$. The expansion of the volume element 
\begin{equation}
\begin{split}
& \frac{\sqrt{-g}}{\sqrt{-g^{(0)}}} = \left[ 1 + \frac{z^2}{2} t^{(2)} + \frac{z^4}{2} \left(t^{(4)} - \frac{1}{2} t^{(2,2)} + \frac{1}{4} (t^{(2)})^2 + \left(\log{z}\right)^2 u^{(0)} + \log{z} \hspace{.1cm} u^{(1)} \right) \right] + \dots \\
& t^{(n)} = \text{Tr}\left[\left(g^{(0)}\right)^{-1} g^{(n)} \right] \hspace{1cm} t^{(2,2)} = \text{Tr}\left[\left(g^{(0)} \right)^{-1} g^{(2)}\left(g^{(0)} \right)^{-1} g^{(2)} \right] \\
& u^{(n)} = \text{Tr}\left[\left(g^{(0)}\right)^{-1} h^{(n)} \right]  \\
\end{split}
\end{equation}
will be needed when expanding the action. The $ij$ component of the Einstein field equation to order $\mathcal{O}(z^{0})$ is solved by
\begin{equation}
g_{ij}^{(2)} = -\frac{1}{2} \left( R[g^{(0)}]_{ij} -\frac{1}{6} R[g^{(0)}] g_{ij}^{(0)} \right)
\end{equation}
which implies
\begin{equation}
\begin{split}
& t^{(2)} = -\frac{1}{6} R[g^{(0)}] \\
& t^{(2,2)} = \frac{1}{4}\left( R[g^{(0)}]_{ij} R[g^{(0)}]^{ij} - \frac{2}{9} R[g^{(0)}]^2 \right)
\end{split}
\end{equation}
The $zz$ component of the Einstein field equation to order $\mathcal{O}(z^{2})$ is solved by
\begin{equation}
\begin{split}
& u^{(0)} = -\frac{2}{3}\left(3b_{1}^2 + c_{1}^2 \right)\\
& u^{(1)} = -\frac{4}{3}\left(3b_{1} b_{2} + c_{1} c_{2} \right) \\
& 4 t^{(4)} = t^{(2,2)} - u^{(0)} - 3 u^{(1)} - \left(3b_{1}^2 + c_{1}^2 \right) - \frac{8}{3} \left(3b_{2}^2 + c_{2}^2 \right) - 4\left(3b_{1}b_{2}+c_{1}c_{2} \right) \\
& \hspace{1.5cm} + \frac{1}{12} \left( |F|_{g^{(0)}}^2 + |G|_{g^{(0)}}^2 \right)
\end{split}
\end{equation}
where $|F|_{g^{(0)}}^{2} = F_{ij} F_{kl} g^{(0)ik} g^{(0)jl} $ is the norm of the boundary field strength and similarly for $|G|_{g^{(0)}}^{2}$.
The leading divergence takes the form
\begin{equation}
\frac{1}{\epsilon^4} \int_{\partial \mathcal{M}} d^{4}x \sqrt{-g^{(0)}} \left(-1 + 4 \right)
\end{equation}
where the coefficients come from $S_{\text{bulk}}$ and $S_{GH}$ respectively. This is cancelled by the counterterm $\delta S_{1} = -3\int_{\partial \mathcal{M}} d^{4}x \sqrt{-h} $. The subleading divergences are
\begin{equation}
\frac{1}{\epsilon^2} \int_{\partial \mathcal{M}} d^{4}x \sqrt{-g^{(0)}} \left(-1 + 1 - \frac{3}{2} \right)t^{(2)}
\end{equation}
where the coefficients come from $S_{\text{bulk}}, S_{GH},$ and $ \delta S_{1}$ respectively. This can be cancelled by the counterterm $\delta S_{2} = -\frac{1}{4} \int_{\partial \mathcal{M}} d^{4}x \sqrt{-h} R[h] $. The logarithmic divergences are given by
\begin{equation}
\begin{split}
& S_{\text{bulk}} \sim \left[ \frac{1}{2} \left( \left(t^{(2)}\right)^2 - t^{(2,2)} \right) -\frac{1}{6} \left(3b_{1}^2 + c_{1}^2 \right) + \frac{1}{8} \left( |F|_{g^{(0)}}^{2} + |f|_{g^{(0)}}^{2} \right)\right] \log{\epsilon} \\
& S_{GH} \sim \frac{2}{3} \left(3b_{1}^2 + c_{1}^2 \right) \log{\epsilon} \\
& \delta S_{1} \sim \left(3b_{1}^2+c_{1}^2\right) \left( \log{\epsilon}\right)^2 + 2\left(3b_{1}b_{2} + c_{1}c_{2} \right) \log{\epsilon} \\
& \delta S_{2} \sim 0 \cdot \log{\epsilon}
\end{split}
\end{equation}
The logarithmic divergences are cancelled by the counterterms
\begin{equation}
\begin{split}
\delta S_{3} =  & \frac{1}{8} \int d^{4}x \sqrt{-h} \log{\epsilon} \left[ \left( R[h]^{ij} R[h]_{ij} - \frac{1}{3}R[h]^2 \right) -  F^{ij} F_{ij} - G^{ij} G_{ij} \right] \\
& + \int d^{4}x \sqrt{-h} \left[ -3(\Sigma-1)^2 -\frac{3}{2\log{\epsilon}} (\Sigma-1)^2  - \phi_{3}^2 - \frac{1}{2\log{\epsilon}} \phi_{3}^2 \right]
\end{split}
\end{equation}
Putting together the different contributions, the renormalized action
\begin{equation}\label{sren}
S_{\text{ren}} = \lim_{\epsilon \rightarrow 0} \left( S_{\text{bulk}} + S_{GH} + \delta S_{1} + \delta S_{2} + \delta S_{3} \right)
\end{equation}
evaluates to 
\begin{equation}
S_{\text{ren}} = \left(\frac{5}{8} - q_{2}^2 \right) \text{Vol}(AdS_{3}) \text{Vol}(S^{1})
\end{equation}
for the surface defect where $\text{Vol}(AdS_{3})$ is the regularized volume of  the $AdS_3$ factor.
\subsection{Vacuum Expectation Values}
Using the renormalized action (\ref{sren}), the vacuum expectation values can be computed through differentiation
\begin{equation}
\begin{split}
& \langle \mathcal{O}_{\Sigma} \rangle = \frac{1}{\sqrt{-g^{(0)}}} \frac{\delta S_{\text{ren}}}{\delta b_{1}} \bigg|_{b_{1}=0} = -3b_{2} \\
& \langle \mathcal{O}_{\phi_{3}} \rangle = \frac{1}{\sqrt{-g^{(0)}}}\frac{\delta S_{\text{ren}}}{\delta c_{1}} \bigg|_{c_{1}=0} = -c_{2} \\
& \langle \mathcal{J}^{i} \rangle = \frac{1}{\sqrt{-g^{(0)}}} \frac{\delta S_{\text{ren}}}{\delta A_{1i}} \bigg|_{A_{1}=0} = \frac{1}{2} \left( A_{3} + 2A_{2} \right)^{i} \\
& \langle j^{i} \rangle = \frac{1}{\sqrt{-g^{(0)}}} \frac{\delta S_{\text{ren}}}{\delta a_{1i}} \bigg|_{a_{1}=0} = \frac{1}{2} \left( a_{3} + 2a_{2} \right)^{i} \\
& \langle T_{ij} \rangle = -\frac{2}{\sqrt{-g^{(0)}}} \frac{\delta S_{\text{ren}}}{\delta g^{(0)ij}} = \lim_{\epsilon \rightarrow 0} \left( \frac{1}{\epsilon^2} T[h]_{ij} \bigg|_{z=\epsilon} \right)
\end{split}
\end{equation}
where $T[h]_{ij}$ is the boundary stress tensor. For the surface defect solution, the asymptotic expansion is
\begin{equation}
r = \frac{1}{z}  + \left( \frac{1}{4} - \frac{q_{2}^2}{3} \right) z - \frac{q_{2}^4}{36} z^3 + \frac{108 q_{2}^2 + 63 q_{2}^4 - 20 q_{2}^6}{3888} z^5 + \dots
\end{equation}
and the expectation values are
\begin{equation}
\begin{split}
& \langle \mathcal{O}_{\Sigma} \rangle = -q_{2}^2 \\
& \langle \mathcal{O}_{\phi_{3}} \rangle = -q_{2} \\
& \langle \mathcal{J}_{\theta} \rangle = q_{2} (1+q_{2}) \\
& \langle j_{\theta} \rangle = q_{2} (1-q_{2}) \\
& \langle T_{ij} \rangle = \left( \frac{3}{8} - 2q_{2}^2 \right)
\begin{pmatrix} 
-\frac{1}{3} g_{AdS_{3}} & 0\\ 
0 & g_{S^{1}}
\end{pmatrix}_{ij}
\end{split}
\end{equation}
so that there are no conformal anomalies: $\langle T_{i}^{i} \rangle = 0$. Note that the solution does not contain any logarithmic divergences and the boundary stress tensor is therefore given by
\begin{equation}
T[h]_{ij} = K_{ij} - K h_{ij} + 3 h_{ij} - \frac{1}{2} \left( R[h]_{ij} - \frac{1}{2} R[h] h_{ij} \right)+ \left( 3(\Sigma-1)^2 + \phi_{3}^2 \right) h_{ij}
\end{equation}

\section{Discussion}
\label{sec5}

In this paper, we investigated solutions of $D=5$, $N=4$ gauged supergravity that are holographic duals of half-BPS conformal surface defects in a $N=2$ SCFT. The ansatz for the solution is informed by the unbroken symmetries of such defects and is given by $AdS_3\times S^1$ warped over an interval with non-trivial gauge potentials along $S^{1}$. We showed for pure Romans' theory that the only solution  in  this class which is non-singular is the $AdS_5$ vacuum; all non-trivial solutions suffer from a conical defect. This situation is improved by coupling vector multiplets to $N=4$ gauged supergravity. The simplest case of one additional vector multiplet already allows for the construction of a one parameter family of regular solutions dual to conformal surface defects preserving eight of the sixteen supersymmetries of the vacuum.

An important question is whether solutions of lower dimensional gauged supergravities can be uplifted to ten or eleven dimensional solutions for which the dual SCFTs are in general known from decoupling limits of brane configurations. It has been shown that pure Romans' theory is a consistent truncation of type IIB \cite {Lu:1999bw,Cvetic:2000yp}, type IIA  \cite{Cvetic:1999xp} and M-theory \cite{Gauntlett:2007sm} and hence solutions of this theory can be uplifted. Much less is known about  uplifts of  matter coupled $D=5, N=4$ gauged supergravity. In \cite{Corrado:2002wx}, it was argued that Romans' theory coupled to two tensor multiplets is a consistent truncation of an orbifold of $AdS_5\times S^5$.  Recently, in  \cite{Cheung:2019pge,Cassani:2019vcl} a consistent truncation of 11-dimensional supergravity on Maldacena-Nunez geometries was constructed, leading to D=5, N=4 gauged supergravity including three vector multiplets.

The rigidity of supersymmetric $N=4$ vacua \cite{Louis:2015dca} makes the existence of other consistent truncations likely.  

Since our  solution has only two gauge fields and scalars turned on, it can be related to solutions in  $D=5,N=2$ gauged supergravity \cite{Behrndt:1998jd,Behrndt:1998ns}. It has been shown in \cite{Cvetic:1999xp} that these solutions can be uplifted to ten and  eleven dimensions, which means that our solution can be uplifted too. It was argued in \cite{Corrado:2002wx} that the truncation used in our paper fall into a class of truncations of gauged  $N=8$ supergravity which can be uplifted to ten dimensions \cite{Khavaev:1998fb}.  
 One could also consider applying the construction in our paper to general class of the gauged supergravities of \cite{Corrado:2002wx} which describe $Z_N$ orbifolds and investigate whether in the field theory, the surface operators of the orbifold theory  can be obtained from surface operators of  $N=4$ SYM \cite{Gukov:2006jk,Gomis:2007fi,Drukker:2008wr}.  We leave these interesting questions for future work.

\section*{Acknowledgements}
The work of M.G.~is supported in part by the National Science Foundation under grant PHY-19-14412. 
M.V. is grateful to the Bhaumik Institute for Theoretical Physics for support.

\newpage

\appendix

\section{Conventions and Supersymmetry}
\label{appa}

The frame field for the metric
\begin{equation*}
ds^{2} = r^{2} (H_{1} H_{2})^{1/3} \left(-\cosh^{2}{\rho} \hspace{.1cm} dt^2 + d\rho^2 + \sinh^{2}{\rho} \hspace{.1cm} d\varphi^{2} \right) + \frac{f}{(H_{1}H_{2})^{2/3}} d\theta^{2} + \frac{(H_{1}H_{2})^{1/3}}{f} dr^2
\end{equation*}
is chosen to be
\begin{equation}
\begin{split}
& e^{0} = r (H_{1} H_{2})^{1/6} \cosh{\rho} \hspace{.1cm} dt \hspace{1cm} e^{1} = r (H_{1} H_{2})^{1/6} d\rho  \hspace{1cm} e^{2} = r (H_{1} H_{2})^{1/6}\sinh{\rho} \hspace{.1cm} d\varphi  \\
& e^{3} = \frac{f^{1/2}}{(H_{1} H_{2})^{1/3}} d\theta \hspace{2.2cm} e^{4} = \frac{(H_{1} H_{2})^{1/6}}{f^{1/2}} dr \\
\end{split}
\end{equation}
The spin connection is then given by
\begin{equation}
\begin{split}
& \omega^{01} = \sinh{\rho} \hspace{.1cm} dt \\
& \omega^{04} = \frac{f^{1/2}}{(H_{1}H_{2})^{1/6}} \frac{d}{dr} \left( r(H_{1}H_{2})^{1/6} \right) \cosh{\rho} \hspace{.1cm} dt \\
& \omega^{12} = -\cosh{\rho} \hspace{.1cm} d\varphi \\
& \omega^{14} =  \frac{f^{1/2}}{(H_{1}H_{2})^{1/6}} \frac{d}{dr} \left( r(H_{1}H_{2})^{1/6} \right) \hspace{.1cm} d\rho \\
& \omega^{24} =  \frac{f^{1/2}}{(H_{1}H_{2})^{1/6}} \frac{d}{dr} \left( r(H_{1}H_{2})^{1/6} \right) \sinh{\rho} \hspace{.1cm} d\varphi \\
&\omega^{34} =\frac{f^{1/2}}{(H_{1}H_{2})^{1/6}} \frac{d}{dr} \left( \frac{f^{1/2}}{(H_{1}H_{2})^{1/3}} \right) d\theta
\end{split}
\end{equation}
All fermions satisfy the symplectic Majorana condition
\begin{equation}
\epsilon_{a}^{\ast} = B \Omega_{ab} \epsilon^{b}
\end{equation}
where $B$ is related to the usual charge conjugation matrix $C$ by $B=\gamma_{0} C$. An explicit basis for the spacetime $\gamma$ matrices in the signature $(-,+,+,+,+)$ is
\begin{equation}
\begin{split}
&\gamma_{0} = i \sigma_{1} \otimes \mathbbm{1} \\
&\gamma_{1} = \sigma_{2} \otimes \mathbbm{1} \\
&\gamma_{2} = \sigma_{3} \otimes \sigma_{1} \\
&\gamma_{3}=\sigma_{3} \otimes \sigma_{2} \\
&\gamma_{4}=\sigma_{3}\otimes \sigma_{3} \\
&B =  \mathbbm{1} \otimes \sigma_{2}
\end{split}
\end{equation}
A basis for the Euclidean Clifford algebra $\Gamma$ is
\begin{equation}
\begin{split}
& \Gamma_{1}= \sigma_{1} \otimes \mathbbm{1} \\
& \Gamma_{2} = \sigma_{3} \otimes \sigma_{1} \\
& \Gamma_{3} = \sigma_{3} \otimes \sigma_{3} \\
& \Gamma_{4} = \sigma_{2} \otimes \mathbbm{1} \\
& \Gamma_{5} = \sigma_{3} \otimes \sigma_{2} \\
& \Omega= \sigma_{1} \otimes \sigma_{2}
\end{split}
\end{equation}
In the chosen gauging, 
\begin{equation}
\begin{split}
&\zeta^{ij} = -\frac{1}{2\sqrt{2}} \Sigma^2 \Gamma_{45}^{ij} \\
&\zeta^{aij} = 0 \\
&\rho^{ij} = \frac{1}{2\sqrt{2}} \frac{\cosh{\phi_{3}}}{\Sigma} \Gamma_{45}^{ij} \\
&\rho^{aij} = -\frac{1}{2} \delta^{a}_{1} \frac{\sinh{\phi_{3}}}{\Sigma} \Gamma_{345}^{ij}
\end{split}
\end{equation}
Using the explicit solution to the equations of motion, the dilatino and gaugino variations both lead to the projection condition
\begin{equation}
\left( \Gamma_{45} \right)_{i}^{\hspace{.15cm}j}\epsilon_{j} = \frac{1}{r (H_{1}H_{2})^{1/2}} \left( \gamma_{34} \Gamma_{3} - i \sqrt{f} \gamma_{4} \right)_{i}^{\hspace{.15cm}j}\epsilon_{j}
\end{equation}
Substituting this projector into the AdS$_{3} \times S^{1}$  gravitino variations gives
\begin{equation}
\begin{split}
& \left( \partial_{t} + \frac{1}{2} \sinh{\rho} \hspace{.05cm} \gamma_{01} - \frac{i}{2} \cosh{\rho} \hspace{.05cm} \gamma_{034} \Gamma_{3} \right)_{i}^{\hspace{.15cm}j}\epsilon_{j} = 0 \\
&\left( \partial_{\rho} - \frac{i}{2} \gamma_{134} \Gamma_{3} \right)_{i}^{\hspace{.15cm}j}\epsilon_{j} = 0 \\
& \left(\partial_{\varphi} - \frac{1}{2} \cosh{\rho} \hspace{.05cm} \gamma_{12} - \frac{i}{2} \sinh{\rho} \hspace{.05cm} \gamma_{234} \Gamma_{3} \right)_{i}^{\hspace{.15cm}j}\epsilon_{j} = 0 \\
& \left( \partial_{\theta} - \left( \mu_{3} - \frac{1}{2} \right) \Gamma_{345} \right)_{i}^{\hspace{.15cm}j}\epsilon_{j} = 0
\end{split}
\end{equation}
These equations can be integrated to
\begin{equation}
\begin{split}
\epsilon_{i} =  \text{exp} &\left( \theta \left(\mu_{3} - \frac{1}{2}\right) \Gamma_{345}  \right)_{i}^{\hspace{.15cm}j} \text{exp}\left( \frac{i\rho}{2} \gamma_{134} \Gamma_{3} \right)_{j}^{\hspace{.15cm}k} \\
& \times \text{exp}\left( \frac{it}{2} \gamma_{034} \Gamma_{3} \right)_{k}^{\hspace{.15cm}l} \text{exp}\left( \frac{\varphi}{2} \gamma_{12} \right)_{l}^{\hspace{.15cm}m} \tilde{\epsilon}_{m}(r)
\end{split}
\end{equation}
Anti-periodicity of $\epsilon_{i}$ under $\theta \rightarrow \theta + 2\pi$ requires the chemical potential to be quantized $\mu_{3} \in \mathbb{Z}$.
After multiplying by $\gamma_{34} \Gamma_{3}$, the projection condition can be expressed in the form
\begin{equation}
\left( 1 + i \sqrt{f} \gamma_{3} \Gamma_{3} + r \sqrt{H_{1}H_{2}} \gamma_{34} \Gamma_{345} \right)_{i}^{\hspace{.15cm}j} \epsilon_{j} = 0
\end{equation}
Similarly multiplying by $\Gamma_{45}$ leads to
\begin{equation}
\left( 1 - i \frac{\sqrt{f}}{r\sqrt{H_{1}H_{2}}} \gamma_{4} \Gamma_{45} + \frac{1}{r\sqrt{H_{1}H_{2}}} \gamma_{34} \Gamma_{345} \right)_{i}^{\hspace{.15cm}j} \epsilon_{j} = 0
\end{equation}
Using these equations, the radial gravitino equation can be put into the form
\begin{equation}
\partial_{r} \epsilon_{i} = \left(a + b \gamma_{34} \Gamma_{345} \right) \epsilon_{i}
\end{equation}
The solution to equations of this form \cite{Romans:1991nq} is
\begin{equation}
\begin{split}
\tilde{\epsilon}_{i}(r) = \frac{1}{r(H_{1}H_{2})^{1/6}} & \left( \sqrt{r \sqrt{H_{1}H_{2}} +1} + i \gamma_{4} \Gamma_{45} \sqrt{r \sqrt{H_{1}H_{2}} -1}  \right)_{i}^{\hspace{.15cm} j} \\
& \times \left(1-\gamma_{34}\Gamma_{345} \right)_{j}^{\hspace{.15cm} k} \left(\epsilon_{0} \right)_{k}
\end{split}
\end{equation}
for some constant symplectic Majorana spinor $\epsilon_{0}$. It can be checked explicitly that the above Killing spinor satisfies the symplectic Marjorana condition.

\section{Half-BPS Line Defect Solution}
\label{appb}

A half-BPS solution describing a superconformal line defect can be constructed in the Euclidean version of pure Romans' supergravity. In the notation of \cite{Bobev:2019ylk}, the supersymmetry variations are
\begin{equation}
\begin{split}
& \delta \psi_{\mu} = D_{\mu}\epsilon - \frac{1}{12} \gamma_{\mu} W \hat{\sigma}_{3} \epsilon+ \frac{i}{12} \left( \gamma_{\mu}^{\hspace{.20cm} \nu \rho} - 4 \delta_{\mu}^{\nu} \gamma^{\rho} \right)h_{\nu \rho} \epsilon \\
& \delta \chi = -\frac{i}{2\sqrt{2}} \left( \gamma^{\mu} \partial_{\mu} \lambda + \partial_{\lambda} W \hat{\sigma}_{3} + i \gamma^{\mu \nu} \partial_{\lambda} h_{\mu \nu} \right)\epsilon
\end{split}
\end{equation}
with
\begin{equation}
\begin{split}
& W = 2(2X+X^{-2}) \\
& h_{\mu \nu} = X^{-1} \left( F_{\mu \nu}^{i} \hat{\sigma}_{3} \sigma_{i} + B_{\mu \nu}^{+} \hat{\sigma}_{-} + B_{\mu \nu}^{-} \hat{\sigma}_{+} \right) - i X^{2} f_{\mu \nu} \\
& X = e^{-\lambda/\sqrt{6}}
\end{split}
\end{equation}
The superconformal line defect preserves an $SO(1,2) \times SO(3)$ bosonic symmetery which can be realized by the ansatz
 \begin{equation}
\begin{split}
& ds^{2} = f_{1}(y)^2 ds_{\mathbb{H}^2}^2 + f_{2}(y)^2 d\Omega_{2}^2 + f_{3}(y)^2 dy^2 \\
& B^{-} = C_{1}(y) \text{vol}_{\mathbb{H}^2} + C_{2}(y) \text{vol}_{S^2}\\
\end{split}
\end{equation}
A similar solution containing only these fields was analyzed in \cite{Bobev:2019ylk}. Imposing the projection condition $\hat{\sigma}_{3} \epsilon = \epsilon$, gives
\begin{equation}
\begin{split}
& \delta \psi = D_{\mu} \epsilon - \frac{1}{2} \gamma_{\mu} \epsilon \\
& \delta \chi = 0
\end{split}
\end{equation}
which are the BPS equations describing AdS$_{5}$. Thus the tensor field $B^{-}$ breaks half the supersymmetries and does not backreact on the metric. $C_{1}(y)$ and $C_{2}(y)$ are determined by the tensor field equation of motion
\begin{equation}
dB^{-} + \ast B^{-} = 0
\end{equation}
The full solution is
\begin{equation}
\begin{split}
& f_{1} = \cosh{y} \\
& f_{2} = \sinh{y} \\
& f_{3} = 1 \\
& C_{1} = \frac{a}{\sinh{y}} + b \left( \frac{y}{\sinh{y}} + \cosh{y} \right) \\
& C_{2} = \frac{a}{\cosh{y}} + b \left( \frac{y}{\cosh{y}} - \sinh{y} \right)
\end{split}
\end{equation}
Using the coordinates
\begin{equation}
\begin{split}
& ds_{\mathbb{H}^2}^2 = \frac{d\tau^2 + dx^2}{x^2} \\
& d\Omega_{2}^2 = d\theta^2 + \sin^2{\theta} d\phi^2
\end{split}
\end{equation}
the solution can be mapped to Euclidean Poincar\'e coordinates
\begin{equation}
ds^2 = \frac{1}{z^2} \left( d\tau^2 + dz^2 + dr^2 + r^2 \left(d\theta^2 + \sin^{2} \theta d\phi^2 \right) \right)
\end{equation}
through the coordinate transformation
\begin{equation}
z = \frac{x}{\cosh{y}} \hspace{1cm} r = x \tanh{y}
\end{equation}
In this coordinate system, the tensor field takes the form
\begin{equation}
\begin{split}
& B^{-} = \tilde{C}_{1} d\tau \wedge dr + \tilde{C}_{2} d\tau \wedge dz + \tilde{C_{3}} \sin{\theta} d\theta \wedge d\phi \\
& r^{-1}\tilde{C}_{1} = z^{-1} \tilde{C}_{2} = \frac{1}{\left(r^2+z^2\right)^{3/2}} \left[a \frac{z}{r} + b \left( \frac{z}{r} \sinh^{-1}{\left(\frac{r}{z}\right)} + \frac{\sqrt{r^2+z^2}}{z} \right)\right] \\
& \tilde{C}_{3} = a \frac{z}{\sqrt{r^2+z^2}} + b \left( \frac{z}{\sqrt{r^2+z^2}} \sinh^{-1}{\left(\frac{r}{z}\right)}  - \frac{r}{z} \right)
\end{split}
\end{equation}
and the leading behavior of the tensor field at the boundary is
\begin{equation}
B^{-} \sim \left( \frac{br}{z} + \frac{az}{r} \right) \frac{d\tau \wedge dr}{r^2} + \left( -\frac{br}{z} + \frac{az}{r} \right) \sin{\theta} d\theta \wedge d\phi
\end{equation}
giving the source and vacuum expectation values of the dual $\Delta=3$ operator. Since the spacetime is Euclidean AdS$_{5}$, the dual stress tensor vanishes
\begin{equation}
\langle T_{ij} \rangle = 0
\end{equation}
The solution can be uplifted to type IIB supergravity  or D=11 supergravity  \cite{Lu:1999bw,Gauntlett:2007sm}, but the higher form fields become complex when Wick rotating back to Lorentzian signature.
\newpage

\providecommand{\href}[2]{#2}\begingroup\raggedright\endgroup

\end{document}